\documentclass[12pt]{article}
\usepackage{mathptmx}
\usepackage{iopconf}
\usepackage{graphicx}
\usepackage{bm}
\usepackage{amsfonts}
\usepackage{epic,graphicx}
\usepackage{latexsym}

\def\a{\alpha}

\def\d{\delta}
\def\e{\eta}

\def\l{\lambda}

\def\m{\mu}
\def\n{\nu}
\def\Om{\Omega}
\def\tOm{{\widetilde\Omega}}

\def\o{\omega}
\def\s{\sigma}

\def\th{\theta}

\def\e{\varepsilon}

\def\ba{{\bar a}}
\def\bb{{\bar b}}
\def\bc{{\bar c}}

\def\E{E_{10}}
\def\pa{\partial}

\def\be{\begin{equation}}
\def\ee{\end{equation}}
\def\bea{\begin{eqnarray}}
\def\eea{\end{eqnarray}}
\def\nn{\nonumber}

\def\cD{{\cal D}}

\def\cJ{{\cal J}}

\def\cL{{\cal L}}

\def\cQ{{\cal Q}}
\def\cP{{\cal P}}
\def\cO{{\cal O}}

\def\cV{{\cal V}}

\def\Z{\mathbb{Z}}
\def\R{\mathbb{R}}

\def\neun{{\mathfrak{e}_{9}}}
\def\zehn{{\mathfrak{e}_{10}}}
\def\kzehn{{\mathfrak{ke}_{10}}}
\def\sl{{\mathfrak{sl}(10)}}
\def\gl{{\mathfrak{gl}(10)}}
\def\E{{E_{10}}}
\def\KE{{K(E_{10})}}
\def\EKE{{E_{10}/K(E_{10})}}
\def\Dot{{\,{\bf *}\,}}

\newcommand{\Ref}[1]{(\ref{#1})}

\newcommand{\ft}[2]{{\textstyle {\frac{#1}{#2}} }}

\begin{document}
\title{Eleven dimensional supergravity and
the $E_{10}/K(E_{10})$ $\sigma$-model at low $A_9$ levels
\footnote{Invited contribution to the XXV International Colloquium 
on Group Theoretical Methods in Physics, 2-6 August 2004, Cocoyoc, Mexico
(Plenary talk given by H.~Nicolai).}}

\author{T.~Damour\dag\ and  H.~Nicolai\ddag\ }

\affil{\dag\ I.H.E.S., 35, route de Chartres, F-91440 Bures-sur-Yvette,
France}
\affil{\ddag\ Max-Planck-Institut f\"ur Gravitationsphysik
(Albert-Einstein-Institut),\\ M\"uhlenberg 1, D-14476 Potsdam, Germany}

\beginabstract {Recently, the concept of  a nonlinear $\s$-model over 
a coset space $G/H$ was generalized to the case where the group $G$
is an infinite-dimensional Kac-Moody group, and $H$ its (formal) `maximal 
compact subgroup'. Here, we study in detail the one-dimensional 
(geodesic) $\s$-model with $G = \E$ and $H= \KE$. We re-examine the 
construction of this $\s$-model and its  relation to the bosonic sector 
of eleven-dimensional supergravity, up to height 30, by using a new 
formulation of the equations of motion. Specifically, we make systematic
use of $\KE$-orthonormal local frames, in the sense that we decompose the 
`velocity' on $\E/\KE$ in terms of objects which are representations
of the compact subgroup $\KE$. This new perspective may help in extending 
the correspondence between the  $\E/\KE$ $\s$-model and supergravity 
beyond the level currently checked.}

\endabstract

\section{Introduction}

In this contribution we explain in detail the recent construction of
the $\E/\KE$ non-linear $\s$-model of \cite{DaHeNi02} and its relation
to the bosonic sector of $D=11$ supergravity \cite{CJS}. The present 
approach differs from \cite{DaHeNi02} in some important technical 
aspects, in particular the use of  `$\KE$-orthonormal frames', rather 
than `coordinate frames'. We hope that this new perspective 
will allow one to extend the results of \cite{DaHeNi02} to  higher-level 
$\s$-model degrees of freedom, and to more general spatial dependences 
of the supergravity fields than those previously taken into account. 
For other, and in part competing, approaches to the search for 
symmetries that might underly M theory we refer readers to
\cite{We01,HeJuPa02,EnHoTaWe03,EngHou03,BrGaHe04,Cha04} and references 
therein. An analysis very similar to the present one, but based on 
the decomposition of $\E$ under its $D_9 \equiv SO(9,9)$ subgroup
has been carried out recently in \cite{KleNic04}.

The present construction grew out of an attempt to extend the well known 
Belinskii-Khalatnikov-Lifshitz (BKL) analysis of spacelike (cosmological)
singularities  \cite{BKL} in Einstein's theory (possibly with additional 
massless fields) and to deepen the surprising discovery of a profound 
relation between this analysis and the theory of indefinite Kac Moody 
algebras \cite{DH1,DH2,DHJN,DaHeNi02,DaHeNi03}. As described in detail 
in \cite{DaHeNi03}, one considers a big-bang-like spacetime with an initial
singular spacelike hypersurface `located' at time $t=0$, on which
some (but not all) components of the metric become singular. According
to the BKL hypothesis spatial gradients should become less and less
important in comparison with time derivatives as $t\rightarrow 0$.
This suggests that the resulting theory should be effectively 
describable in terms of a one dimensional reduction. Ref. \cite{DaHeNi03}, 
which generalized previous results by BKL and others, made this idea more
precise by \footnote{We take this opportunity to correct three misprints 
in \cite{DaHeNi03}: in eq.~(6.24), the dilaton contribution should 
appear with a plus sign: $+ ({\lambda_p}/2 )\phi$, whereas it should 
come with a minus sign in eq.~(6.29). The argument of the $\Theta$ function 
in eq.~(6.31) should read $-2 \tilde{m}_{a_1\cdots a_{d-p-1}} (\beta)$.}

(i) proving in full generality that, except for a finite number of them,  
the infinite number of degrees of freedom encoded in the spatially 
inhomogeneous metric, and in other fields, \textit{freeze} in the sense 
that they tend to some finite limits as $t\rightarrow 0$;
 
(ii) showing that the dynamics of the remaining `active' degrees of 
freedom (corresponding to the diagonal components of the metric, and 
to the dilaton(s)) could be asymptotically described in terms of a 
simple `billiard dynamics';

(iii) proving that in many interesting physical theories the `billiard 
table' encoding the dynamics of the active degrees of freedom could 
be identified with the Weyl chamber of some Lorentzian Kac-Moody 
algebra; and

(iv) generalizing the concept of nonlinear $\s$-model on a coset 
space $G/H$ to the case where $G$ is a Lorentzian Kac-Moody group, and
$H$ its `maximal compact subgroup', and showing that such (one-dimensional) 
$\s$-models are asymptotically (as $t\rightarrow 0$) equivalent to 
the billiard dynamics describing the active degrees of freedom.

The \textit{correspondence} between `cosmological billiards' and 
`Kac-Moody $\s$-model billiards', \textit{i.e.} geodesics on Kac-Moody 
cosets  $G/H$, is relatively easy to establish when one considers only
the \textit{leading} terms in the dynamics near $t\rightarrow 0$. Going 
beyond this leading behaviour is much harder and has been attempted 
only for pure gravity (in which case the relevant Kac Moody algebra
is $AE_3$ \cite{DHJN,DaHeNi03}), and for the bosonic sector of $D=11$ 
supergravity. We recall that this model includes, besides the metric field 
$g_{\m \n}(t, {\bf x})$, a 3-form $A_{\m \n \l}(t, {\bf x})$ having a 
specific Chern-Simons self-coupling $AFF$, where $F=dA$ \cite{CJS}. 
Ref. \cite{DaHeNi02} 
introduced a precise identification between the purely $t$-dependent 
$\s$-model quantities obtained from the geodesic action on the $\E/\KE$ 
coset space on the one hand, and certain fields of $D=11$ supergravity 
and their spatial gradients evaluated {\em at a given, but arbitrarily 
chosen spatial point} ${\bf x} ={\bf x}_0$ on the other. So far, 
this correspondence works for suitably truncated versions of both 
models, with $\ell\leq 3$ and height $\leq 29$ on the $\s$-model side, 
and zeroth and first order spatial gradients on the supergravity side. 
There are, however, indications that it extends to higher levels 
and higher order spatial gradients: as shown in \cite{DaHeNi02}, 
the level expansion of $\E$ contains all the representations needed 
for the higher order spatial gradients (as well as many other 
representations, see \cite{FisNic03}). This observation gave rise to 
the key conjecture of \cite{DaHeNi02}, according to which the full 
geometrical data of $D=11$ supergravity, or some theory containing it,
can be mapped onto a geodesic motion in the $\E/\KE$ coset space.
The hope %(which has been verified so far only up to the Kac-Moody height 30)
is that the infinite number of degrees of freedom associated to the 
ten-dimensional spatial gradients of the metric and the 3-form (and 
possibly to other M-theoretic degrees of freedom) can be put in 
one-to-one correspondence with the infinite number of parameters of the 
$\E/\KE$ coset space \footnote{The residual {\em spatial dependence}, 
which on the $\s$-model side is supposed `to be spread' over the 
$\E$ Lie algebra, is the main difference of the present scheme with a 
{\it bona fide} reduction of $D=11$ supergravity to one dimension, 
for which the appearance of $\E$ had first been conjectured \cite{JE10}.}. 
Another way to view this proposal is in terms of a `small tension expansion' 
of the full theory, which in turn might be related to the zero tension 
limit of string theory \cite{Gross}.

We emphasize that no extra assumptions beyond the geodesic action are
required in the present setup, and that our proposed geodesic action
is  {\em essentially unique}, as we will show here. In particular, the
relative normalization of all terms in the equations of motion follow 
from that action, independently of the existence of a supersymmetric
extension. For instance, the unique value of the coefficient of 
the Chern-Simons coupling $\propto AFF$ present in the 11-dimensional 
supergravity action \cite{CJS} is found to match exactly a coefficient 
in the $\E/\KE$ geodesic action. As further evidence of the correspondence 
between the two actions one might count the fact that the geodesic 
action on $\E/\KE$ is not compatible with the addition of a cosmological 
constant in the $D=11$ supergravity action (an addition which has 
been proven to be incompatible with supersymmetry in \cite{BDHS}). 
Indeed, it is easily checked that  such a term in the supergravity 
Hamiltonian cannot be matched to any $\E$ root.

In the present paper, we shall introduce a new formulation of the
$\E/\KE$ $\s$-model. In our previous work \cite{DaHeNi02} we had
studied the dynamics defined by the $\E/\KE$ action in terms of what 
were, essentially, some global coordinates $A^{(\ell)}$ on the coset 
space $\E/\KE$. These coordinates are defined by considering a 
Borel-type triangular exponential parametrization of a general 
coset element in the form
\be
\cV\big(A(t)\big) =
\exp \Big( \sum_{\ell =0}^\infty A^{(\ell)}(t) \Dot E^{(\ell)}\Big)
\ee
The notation here is very schematic and will be further explained below. 
Suffice it to say here that  $E^{(0)}$ denote the Cartan and positive-root 
generators of the $GL(10)$ subalgebra of $\E$, while $E^{(\ell)}$, $\ell >0$ 
denote all the remaining raising (positive root) generators of $\E$. 
Here $\ell$ denotes the $GL(10)$ level, and all the $GL(10)$ (and 
degeneracy) indices needed to specify the representations appearing 
among the $E^{(\ell)}$'s are suppressed. The infinite sequence of real 
numbers $A^{(\ell)}$ for $\ell \geq 0$ explicitly defines the specific 
triangular coordinates of any coset class $\cV \in \E/\KE$.
The coordinates $A^{(\ell)}$ are globally defined on $\E/\KE$, and they
were explicitly used in \cite{DaHeNi02} to write the $\s$-model action 
in the second order coordinate form $ S = \int dt \, n^{-1} \cL(A,\pa_t A)$. 
This form leads to Euler-Lagrange equations of motion which contain 
two time derivatives of the coordinates $A^{(\ell)}(t)$.

By contrast  to this second order explicit coordinate form,
$\ddot A = F(A, \dot A)$, we shall work  in this paper with a (formal) 
first-order form, $\dot \cP = F(\cP,\cQ)$ based on what one might 
call `local $\KE$-orthonormal frames', by analogy with the geodesic motion 
on, say, the Lobachevskii plane viewed as the coset $SL(2)/SO(2)$).
Indeed, the objects $\cP$ and $\cQ$ will be defined by decomposing the 
`velocity' $ \cV^{-1} \pa_t \cV$ on $\E$ in terms of objects which are 
representations of the subgroup $\KE$. In the  Lobachevskii plane analogy, 
one would say that the quantities $\cP$ and $\cQ$ carry `flat indices' 
($SO(2)$ indices). In most of our developments, we shall not need
any explicit representation of $\cP$ and $\cQ$ in terms of local 
coordinates on $\E/\KE$. Note, however, that, in any choice of 
parametrization of $\E/\KE$, such as the triangular one mentioned above, 
$\cP$ and $\cQ$ become some functions of $A$ and $\dot A$.

We hope that this new perspective can help in extending the correspondence  
between the  $\E/\KE$ $\s$-model and supergravity beyond the level
where it is currently checked. Indeed, as in \cite{DaHeNi02}, we shall be 
able here to verify this correspondence up to height 29, included. 
Beyond this height, there appear terms in both versions of the equations 
of motion that we will exhibit explicitly, but that we do not know 
how to match. The more streamlined form of the equations of motion 
used here might help in guessing how to extend the `dictionary' 
between the supergravity variables and the $\E/\KE$ ones. In \cite{DaHeNi02} 
some guesses were proposed to relate higher order spatial gradients of 
supergravity variables and higher-level coset variables $A^{(\ell)}$. 
One needs to make these guesses fully concrete, and to check that, together 
with a suitably extended  dictionary, they extend the match between the 
two actions, to confirm the conjecture that $\E$ is a hidden symmetry 
of 11-dimensional supergravity (and M-theory). In a separate 
publication \cite{DNR4}, we shall report some recent progress in 
this direction based on the consideration of the higher order (in 
Planck length) corrections to the low-energy supergravity Lagrangian.

Our use of flat indices here also paves the way for the introduction of
fermionic couplings. The fermions of the theory will transform under
{\em local} $\KE$, which in a supersymmetric extension of the $\E/\KE$ coset 
model would become the `R-symmetry'. $\KE$ contains the spatial Lorentz 
group $SO(10)$ at level zero, and this symmetry is only manifest when 
one formulates the equations of motion entirely in terms of $\KE$ objects 
on the $\s$-model side. The use of flat indices was also found to be
more convenient for the $D_9$ decomposition of $\E$ in \cite{KleNic04},
where the present analysis was extended to fermionic degrees of freedom 
and the compatibility of a Romans type mass term (for IIA supergravity) 
with $\E$ was established.

\section{Basic facts about $E_{10}$}

\subsection{Basic definitions}
The only known way to define the Kac Moody (KM) Lie algebra 
$\zehn \equiv {\rm Lie}\, (\E)$ is via its Chevalley-Serre presentation 
in terms of generators and relations and its Dynkin diagram \cite{Kac,MP}, 
which we give below with our labeling conventions for the simple roots
$\{ \a_i \,|\, i=0,1,...,9\}$:

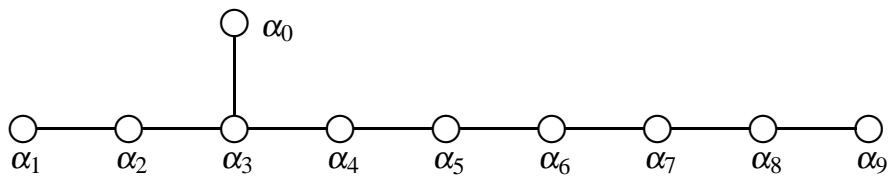
\begin{center}
\scalebox{1}{
\begin{picture}(340,60)
\put(5,-5){$\alpha_1$}
\put(45,-5){$\alpha_2$}
\put(85,-5){$\alpha_3$}
\put(125,-5){$\alpha_4$}
\put(165,-5){$\alpha_5$}
\put(205,-5){$\alpha_6$}
\put(245,-5){$\alpha_7$}
\put(285,-5){$\alpha_8$}
\put(325,-5){$\alpha_9$}
\put(100,45){$\alpha_{0}$}
\thicklines
\multiput(10,10)(40,0){9}{\circle{10}}
\multiput(15,10)(40,0){8}{\line(1,0){30}}
\put(90,50){\circle{10}} \put(90,15){\line(0,1){30}}
\end{picture}}\end{center}
\vspace*{0.4cm}
The nine simple roots $\a_1,\dots,\a_9$ along the horizontal line generate
an $A_9 \equiv \sl$ subalgebra of $\zehn$ \footnote{We will only be 
concerned with these Lie algebras as algebras over the real numbers, 
{\it i.e.} $\sl \equiv \mathfrak{sl}(10,\R)$ and $\zehn\equiv \zehn(\R)$, 
{\it etc.}}. The simple root $\a_0$, which connects to $\a_3$ will be referred 
to as the `exceptional' simple root. Its dual Cartan subalgebra (CSA) 
element $h_0$ enlarges $\sl$ to the Lie algebra $\gl$.

The Lie algebra $\zehn$ is built in terms of multiple commutators of a 
set of basic triples $\{e_i,f_i,h_i\}$, where $i,j=0,1,...9$, and each
triple generates an $A_1\equiv\mathfrak{sl}(2)$ subalgebra. The CSA
is spanned by the generators $\{ h_i\}$, i.e. $[h_i, h_j] = 0$;
the remaining bilinear relations are
\be\label{CS1}
[h_i , e_j ] =  A_{ij} e_j \;\; , \quad 
[h_i , f_j ] =  - A_{ij} f_j \;\; , \quad
[e_i , f_j]  =  \d_{ij} h_i \;\; , \quad 
\ee
where $A_{ij}$ is the $\E$ Cartan matrix. In addition, we have the 
multilinear Serre relations
\be\label{CS2}
({\rm ad} \, e_i)^{1-A_{ij}} (e_j) = 0 \;\;\; , \quad
({\rm ad} \, f_i)^{1-A_{ij}} (f_j) = 0
%[e_i [ \dots [e_i, e_j]...]] = 0 \;\;\; , \quad
%f_i [ \dots [f_i, f_j]...]] = 0
\ee
We will also need the standard bilinear form 
\be\label{Bil}
\langle e_i | f_j \rangle = \d_{ij} \;\; , \quad
\langle h_i | h_j \rangle = A_{ij}
\ee
It extends to the full Lie algebra $\zehn$ by its invariance 
property $\langle [x,y]|z\rangle = \langle x|[y,z]\rangle$.

An important role will be played by the `maximal compact subalgebra' 
${\rm Lie}\, (\KE) =: \kzehn\subset\zehn$. It is defined as the invariant 
Lie subalgebra of $\zehn$ under the Chevalley involution
\be
\th (h_i) = - h_i \;\;\;, \quad
\th (e_i) = - f_i \;\;\;, \quad
\th (f_i) = - e_i 
\ee
This involution extends to all of the Kac Moody Lie algebra by means of 
$\th([x,y]):= [\th (x), \th (y)]$. The associated $\th$-invariant `maximal 
compact subgroup' will be designated by $\KE$ (we put quotation marks because 
$\KE$ is not necessarily compact in the topological sense). It is not 
difficult to see that $\kzehn$  is generated by all multiple commutators 
of the elements $(e_i - f_i)$. The corresponding real form of $\zehn$ is 
the analog of the split forms $E_{n(n)}$ for $n\leq 8$, which is the 
reason for sometimes denoting $\E$ as $E_{10(10)}$. For later use we 
also introduce the notion of the `transposed element': for any element 
$x\in\zehn$ we define
\be\label{T}
x^T := - \th (x)
\ee
In this sense $\kzehn$ consists of all `antisymmetric' elements
$x= -x^T$ of $\zehn$, in the same way that $\mathfrak{so}(10)$ consists 
of all antisymmetric matrices in $\mathfrak{sl}(10)$.

\subsection{Level decomposition w.r.t. $\sl\subset\zehn$}

Because no closed form construction exists for the Lie algebra elements 
$x\in\zehn$, nor their invariant scalar products, we will rely on a 
recursive approach based on the decomposition of $\zehn$ into irreducible 
representations of its $\sl$ subalgebra \footnote{A similar analysis of 
$\zehn$ in terms of its $D_9 \equiv \mathfrak{so}(9,9)$ subalgebra is 
given in \cite{KleNic04}. The decomposition of $\zehn$ under its affine 
$\mathfrak{e}_9$ subalgebra had already been studied in \cite{KMW}.}. 
Any positive root of $\E$ can be written as
\be\label{E10root}
\a = \ell \a_0 + \sum_{j=1}^9 m^j \a_j %\quad (\ell,m^j \geq 0)
\ee
with $\ell, m^j \geq 0$. The integer $\ell \equiv \ell (\a)$ is called
the `$A_9$ level', or simply the `level' of the root $\a$; below, we
will, however, switch conventions by associating positive levels 
with multiple commutators of $f$'s, {\it i.e.} negative roots. The
decomposition \Ref{E10root} corresponds to a slicing (or `grading') 
of the forward lightcone in the root lattice by spacelike hyperplanes, 
with only finitely many roots in each slice (slicings by lightlike 
or timelike hyperplanes would produce gradings w.r.t. affine or 
indefinite KM subalgebras, with each slice containing infinitely 
many roots). Every positive root $\a$ is associated with a set of 
`raising operators' $E_{\a,s}$, where $s=1,...,{\rm mult}\, \a$ counts the 
number of independent such elements of $\zehn$, and ${\rm mult}\, \a$ 
is the `multiplicity' of the root in question; similarly, the 
`lowering operators' are associated with negative roots.

The adjoint action of the $\sl$ subalgebra leaves the level $\ell(\a)$ 
invariant. The set of $\zehn$ elements corresponding to a given level 
$\ell$ can therefore be decomposed into a (finite) number of irreducible 
representations of $\sl$. Because of the recursive definition of $\zehn$
in terms of multiple commutators, all representations occurring at level 
$\ell +1$ are contained in the product of the level-$\ell$ representations 
with the $\ell=1$ representation. The multiplicity of $\a$ as a root of 
$\zehn$ is equal to the sum of its multiplicities as a weight 
occurring in the $\sl$ representations. Each irreducible representation 
of $A_9$ can be characterized by its highest weight $\Lambda$, or
equivalently by its Dynkin labels $(p_1,\dots,p_9)$ where
$p_k := (\a_k,\Lambda)\geq 0$ is the number of columns 
with $k$ boxes in the associated Young tableau. For instance, the
Dynkin labels $(001 000 000)$ correspond to a Young tableau consisting 
of one column with three boxes, {\it i.e.} the antisymmetric tensor 
with three indices. The Dynkin labels are related to the 9-tuple of 
integers $(m^1,\dots,m^9)$ appearing in \Ref{E10root} (for the
highest weight $\Lambda\equiv - \a$) by
\be\label{mi}
 S^{i3} \ell - \sum_{j=1}^9 S^{ij} p_j = m^i \geq 0
\ee
where $S^{ij}$ is the inverse Cartan matrix of $A_9$. This relation 
strongly constrains the representations that can appear at level $\ell$,
because the entries of $S^{ij}$ are all positive, and the 9-tuples 
$(p_1,\dots,p_9)$ and $(m^1,\dots, m^9)$ must both consist of 
non-negative integers. In addition to satisfying the Diophantine 
equations \Ref{mi}, the highest weights must be roots of $\E$, which 
implies the inequality 
\be\label{L2}
\Lambda^2 = \a^2 =
 \sum_{i,j=1}^9 p_i S^{ij} p_j - \ft1{10} \ell^2 \leq 2
\ee

The problem of finding an explicit representation of $\zehn$ 
in terms of an infinite tower of $\sl$ representations can thus be 
reformulated as the problem of identifying all $\sl$ representations
compatible with the Diophantine inequalities \Ref{mi}, \Ref{L2}. 
The more difficult task is to determine their outer multiplicities, 
{\it i.e.} the number of times each representation appears at a given 
level $\ell$. Making use of the known root multiplicities of $\zehn$ 
it is possible to determine the level decomposition and the outer 
multiplicities of all representations to rather high levels (up to 
$\ell =28$ so far \cite{FisNic03}; analogous tables for the $D_9$ 
decomposition of $\zehn$ are given in \cite{KleNic04}).

Let us now describe the lowest levels of this decomposition in detail
\footnote{Modulo dimensions, the representations at the first three 
levels are actually the same for all $E_{n+1}$ in the decomposition 
w.r.t. $A_n$, see \cite{We01} for information concerning $E_{11}$.}. 
For $E_{10}$, the level $\ell=0$ sector is just the $\gl$ subalgebra 
spanned by $A_9$ and the exceptional CSA generator $h_0$. 
In `physicists' notation', this algebra is written as
\be
[{K^a}_b , {K^c}_d ] =  \d^c_b {K^a}_d -\d^a_d  {K^c}_b  
\ee
with indices $a,b,...\in\{ 1,...,10\}$. Note that $({K^a}_b)^T = {K^b}_a$.

The level $\ell=1$ elements transform in the $(001000000)$
representation of $\sl$, i.e. as a 3-form; they are thus 
represented by the $\gl$ tensor $E^{abc}$. The Chevalley 
conjugate elements at level $\ell =-1$ are 
\be\label{adjoint}
F_{abc} = \big( E^{abc} \big)^T  
\ee
and transform in the contragedient representation; thus
\be
[{K^a}_b , E^{cde} ] =  3 \d^{[c}_b E^{de]a} \;\; , \quad
[{K^a}_b , F_{cde} ] =  - 3 \d^{a}_{[c} F_{de]b}
\ee
The remaining level $|\ell| \leq \pm 1$ commutators are 
\be
[F_{abc}, E^{def}] 
 = -18 \, \d^{[de}_{[ab} {K^{f]}}_{c]} + 2\, \d^{def}_{abc} K
\ee
where $K:= {K^a}_a$. We normalize all antisymmetric objects
with weigth one, so that, say, $A_{[ab]} = \ft12 (A_{ab} -A_{ba})$,
and $\d^{def}_{abc} = \ft16 (\d^d_a \d^e_b \d^f_c + 5 \; {\rm terms})$.
Hence, the normalization of the last equation is such that, e.g.
$$
[F_{123},E^{123}] = -({K^1}_1 + {K^2}_2 + {K^3}_3) + \ft13 K 
$$
The above elements are already sufficient to identify the Chevalley 
generators of $\zehn$: we have
\be
e_0 = F_{123} \;\; ,  \quad f_0 = E^{123}  \;\; ,  \quad 
   h_0 = - {K^1}_1 - {K^2}_2 - {K^3}_3  + \ft13 K     
\ee
for the exceptional node, and
\be
e_i = {K^i}_{i+1} \;\; ,  \quad f_i = {K^{i+1}}_i \;\; ,  \quad 
       h_i = {K^i}_i - {K^{i+1}}_{i+1}
\ee
for the remaining nodes with $1\leq i \leq 9$ which generate the
the $\sl$ subalgebra. With the scalar products
\be
\langle {K^a}_b | {K^c}_d \rangle = \d^a_d \d^c_b - \d^a_b \d^c_d \quad ,
\qquad \langle F_{abc} | E^{def} \rangle = 3! \, \d_{abc}^{def}
\ee
it is straightforward to recover the bilinear form (\ref{Bil}) above.

There are two elements of the CSA which play a distinguished role:
the {\it central charge} $c$ of the affine subalgebra 
$\neun\subset\zehn$ is given by
\be
c = 2h_1  + 4 h_2 + 6 h_3 + 5 h_4 + 4 h_5 + 3 h_6 + 2 h_7 + h_8 + 3 h_0
  = {K^{10}}_{10}
\ee
The affine subalgebra $\neun$ must commute with the central charge;
its Chevalley generators are obtained from the above set by omitting
the triple $\{e_9,f_9,h_9\}$. The affine algebra is thus generated from 
the level $|\ell|\leq 1$ elements by restricting the indices 
$a,b,...$ to the values $\in \{1,...,9\}$. The affine level counting
(alias mode counting) operator is
\be 
d= c + h_9 = {K^9}_9
\ee
It tells us that the affine mode number of a given affine element
is equal to the difference of the number of upper and 
lower indices equal to $9$  \footnote{In the $d=2$ reduction, 
where one is left with the dependence on the time and one space 
coordinate $x^a\equiv x^{10}$, and maximal supergravity is known to 
admit an affine $E_9$ symmetry, these generators are realized as follows: 
$c$ acts on the conformal factor \cite{Julia,BM}, whereas $d$ acts as
a dilatation operator \cite{BW,Maison,JN}.}.

Similarly to the decomposition of $\zehn$ in terms of $\sl$ representations,
we can analyze its invariant subalgebra $\kzehn$ in terms of its
$\mathfrak{so}(10)$ subalgebra, the invariant subalgebra of $\sl$.
At lowest order, we have
\be\label{Lab}
L_{ab} := \ft12 \big( {K^a}_b - ({K^a}_b)^T\big) 
   \equiv \ft12 ({K^a}_b - {K^b}_a) 
\;\;\; , \quad L_{abc} := \ft12 (E^{abc} - F_{abc})
\ee
Note that the $\kzehn$ elements combine level $\ell$ with level
$-\ell$ elements. For them, the (upper or lower) position of indices 
no longer matters, as they have to be regarded as $SO(10) \equiv K(GL(10))$ 
rather than $GL(10)$ indices. We will also make use of the coset space 
generators
\be\label{Sab}
S_{ab} := \ft12 \big( {K^a}_b + ({K^a}_b)^T\big) 
 \equiv \ft12 ({K^a}_b + {K^b}_a )
\;\;\; , \quad S_{abc} := \ft12 (E^{abc} + F_{abc})
\ee

\subsection{Levels $\ell= \pm 2 , \pm 3$}

At levels 2 and 3 we have the representations $(000001000)$ and
$(100000010)$, which are respectively generated by
\be\label{level23}
E^{a_1 \dots a_6} := [ E^{a_1 a_2 a_3} ,  E^{a_4 a_5 a_6}] \quad , \quad
E^{[ a_0 | a_1 a_2] a_3 \dots a_8} := [ E^{a_0 a_1 a_2}, E^{a_3 \dots a_8}]
\ee
Making use of $E^{a_0 | a_1 \dots a_8} = -8 E^{[a_1 | a_2 \dots a_8] a_0}$
the last relation can be rewritten in the form
\be
E^{a_0|a_1\dots a_8} = 4 \, [ E^{a_0 [a_1 a_2}, E^{a_3 \dots a_8]}] 
\ee
The adjoint ($\ell=-2,-3$) elements are
\bea
F_{a_1 \dots a_6} &:=& \big( E^{a_1 \dots a_6} \big)^T = 
- [ F_{a_1 a_2 a_3} ,  F_{a_4 a_5 a_6}] 
\\
F_{a_0|a_1\dots a_8} &:=&  \big( E^{a_0 | a_1\dots a_8} \big)^T =  
 - 4 \, [ F_{a_0 [a_1 a_2}, F_{a_3 \dots a_8]}] 
\eea
Further commutation yields 
\be
[F_{a_1 a_2 a_3} ,  E^{b_1 \dots b_6} ] 
=  5! \, \d^{[b_1 b_2 b_3}_{a_1 a_2 a_3} E^{b_4 b_5 b_6]} 
\ee
\be
[F_{a_1\dots a_6} , E^{b_1 \dots b_6} ] 
 =  -6\cdot 6! \, \d^{[b_1 \dots b_5}_{[a_1 \dots a_5} {K^{b_6 ]}}_{a_6 ]}
   + \ft23 \cdot 6! \, \d^{b_1 \dots b_6}_{a_1\dots a_6} K
\ee
and
\bea
[ F_{a_1 a_2 a_3} , E^{b_0| b_1 \dots b_8} ]
&=& 7 \cdot 48 \; \Big( \d_{a_1 a_2 a_3}^{b_0 [b_1 b_2} E^{b_3\dots b_8]} 
       - \d_{a_1 a_2 a_3}^{[b_1 b_2 b_3} E^{b_4\dots b_8] b_0} \Big)\nn\\
{}[ F_{a_1 \dots a_6} , E^{b_0| b_1 \dots b_8} ]
&=&  8! \;
\Big( \d_{a_1 a_2\dots \, a_6}^{b_0 [b_1 \dots b_5} E^{b_6 b_7 b_8]}
 - \d_{a_1 \dots \, a_6}^{[b_1 \dots \, b_6} E^{b_7b_8] b_0} \Big) 
\eea
The conjugate relations are easily obtained by taking the transpose 
of these commutators (not forgetting minus signs). The above relations
can be conveniently restated after multiplication of the level $\ell=3$ 
generators by a dummy tensor $X_{a|b_0\dots b_8}$, which gives (after 
some reshuffling of indices by means of Schouten's identity)
\bea
\big[ F_{cde} \, ,\, X_{a|b_1\dots b_8} E^{a|b_1\dots b_8}\big] &=&
7\cdot 72 \; X_{[c|de]b_1 \dots b_6} E^{b_1 \dots b_6} \nn\\
\big[ F_{c_1\dots c_6}\, ,\, X_{a|b_1\dots b_8} E^{a|b_1\dots b_8}\big] &=&
3\cdot 8!\; X_{[c_1|c_2 \dots c_6]def} E^{def} 
\eea
The remaining commutation relation between levels $\ell=3$ and $\ell = -3$ 
is also most easily written in this way
\bea
\big[ F_{c|d_1\dots d_8} \, , \,X_{a|b_1\dots b_8} E^{a|b_1\dots b_8}\big] &=&
%&& \!\!\!\!\!\!\!\!\!\!\!\!\!\!\!\!\!\!\!\!\!\!\!\!\!\!\!
%\!\!\!\!\!\!\!\!\!\!\!\!\!\!\!\!\!\!\!\!\!\!\!\!\!\!\!\!\!\!\!\!\!
8\cdot 9! \; \Big( - X_{c|e[d_1 \dots d_7} {K^e}_{d_8]} -
     \ft19 X_{c|d_1 \dots d_8} K  \nn\\ 
&& - X_{[d_1|d_2 \dots d_8]e} {K^e}_{c}+ \ft19 X_{[d_1|d_2 \dots d_8]c} K \Big)
\eea
The standard bilinear form on these elements is evaluated by making
use of the invariance property $\langle [x,y]|z\rangle =
\langle x| [y,z]\rangle$, with the result
\bea
\langle F_{a_1 \dots a_6} | E^{b_1 \dots b_6} \rangle &=&
     6! \, \d_{a_1 \dots a_6}^{b_1 \dots b_6} \nn\\
\langle F_{a_0|a_1 \dots a_8} | E^{b_0|b_1 \dots b_8} \rangle &=&
     8 \cdot 8! \Big(\d^{a_0}_{b_0} \d^{a_1\dots a_8}_{b_1\dots b_8} -
            \d^{[a_1}_{b_0} \d^{a_2\dots a_8] a_0}_{b_1\dots b_7 b_8} \Big)
\eea
such that e.g. $\langle F_{123456}|E^{123456}\rangle = 1$ and
$\langle F_{1|1 \dots 8} | E^{1|1 \dots 8} \rangle = 9$.

\section{The $\EKE$ $\s$-model for $\ell\leq 3$}

\subsection{General remarks}
In this section we will set up the general formalism for
$\s$-models in one (time) dimension. The geodesic Lagrangian $\cL$
on $\E/K(\E)$ is defined by generalizing the standard Lagrangian 
on a finite dimensional coset space $G/K(G)$, where $K(G)$ is the 
maximal compact subgroup of the Lie group $G$ (for a given real
form of $G$). Despite the formal replacement of the finite dimensional 
groups $G$ and $K(G)$ by the infinite dimensional groups $\E$ and 
$\KE$, all elements entering the construction of $\cL$ have natural 
generalizations to the case where $G$ is the group obtained by 
(formal) exponentiation of an indefinite or hyperbolic KM algebra. 
In particular, our expansion in terms of levels provides us with an 
algorithmic scheme which is completely well defined and computable 
to any given finite order, and which in principle can be carried to 
arbitrarily high levels. An essential ingredient in this construction 
is the so-called triangular gauge, which we will explain below.

One important difference between the finite dimensional coset
spaces $G/K(G)$ and the infinite dimensional space $\EKE$ is
the following. For $K(G)$ the maximal compact subgroup of $G$,
the space $G/K(G)$ is always Euclidean (i.e. endowed with a positive
definite metric). This not so for the space $\EKE$: even though $\KE$
is `compact' in the algebraic sense, the metric on $\EKE$ has 
precisely {\em one negative eigenvalue} coming from the negative
norm CSA generator. It is for this reason that we can define
null (= lightlike) geodesics on $\EKE$ which do not exist in 
the finite dimensional case.

\subsection{$\s$-model and level expansions}
Following the standard formulation of nonlinear $\s$-models for 
$G/K(G)$ coset spaces  with $K(G)$ the maximal compact subgroup of $G$,
we assume the bosonic degrees of freedom are described by a `matrix'
$\cV\in\E$, which itself can be parametrized in terms of coordinates
(`fields') $A^{(\ell)} = A^{(\ell)}(t)$ depending on the affine 
parameter $t$, the time coordinate, for all $\ell\in\Z$. Being an 
element of the coset space $G/K(G)$, the `matrix' $\cV$ is subject 
to rigid and local transformations acting from the left and the 
right, respectively:
\be\label{gVk}
\cV\big(A(t)\big) \longrightarrow g \cV\big(A(t)\big) k(t)  
\quad {\rm with} \;\;\;  g\in\E\;\; , \;\;      k(t) \in K(\E)
\ee
The local $\KE$ invariance allows us to choose a convenient gauge.
For our calculations we will always adopt the triangular gauge where 
by definition all fields corresponding to negative levels are set 
to zero \footnote{By abuse of language, we will use this terminology
even if $\cV$ is triangular only with regard to levels $\ell\geq 1$, 
but not necessarily for $\ell = 0$.}. In other words, in triangular 
gauge $\cV$ is (formally) obtained by exponentiating the Borel 
subalgebra consisting of the level $\ell\geq 0$ elements of $\zehn$, viz.
\be\label{triangular}
\cV\big(A(t)\big) = 
\exp \Big( \sum_{\ell =0}^\infty A^{(\ell)}(t) \Dot E^{(\ell)}\Big)
\ee 
The notation here is slightly schematic: the symbol `$\Dot$' includes a sum 
over all the irreducible representations appearing at level $\ell$ (whose 
number grows very rapidly with $\ell$, see \cite{FisNic03}) as well as all 
indices labeling a particular representation (we will be more specific 
below). With this choice of gauge -- as well as in any other gauge -- 
there remains only the rigid $\E$ symmetry, which is now  realized 
{\em non-linearly}: those $\E$ transformations in \Ref{gVk} which
violate the chosen gauge must be compensated by field dependent
$\KE$ transformations.

In order to write down the geodesic action, we decompose the
Lie algebra valued `velocity' $v \equiv \cV^{-1} \pa_t \cV$ into its
symmetric and antisymmetric parts, {\it i.e.} $\cP_t := \ft12 (v + v^T)$, 
and $\cQ_t := \ft12 (v - v^T)$, respectively. Thus we write
\be\label{VdotV}
\cV^{-1} \pa_t \cV = \cQ_t + \cP_t \in\zehn \quad , \qquad
\cQ_t^T = - \cQ_t \;\; , \;\; \cP_t^T = + \cP_t
\ee
(see \Ref{T} for the definition of `transposition'). Hence $\cQ_t\in \kzehn$,
and $\cP_t$ belongs to the coset $\zehn \ominus \kzehn$. For convenience
of notation we will omit the subscript $t$ in the remainder, such that
$\cQ \equiv \cQ_t$ and $\cP \equiv \cP_t$
will be understood to be `world tensors' under reparametrizations of the 
time coordinate $t$. From (\ref{gVk}) it follows immediately that the 
quantities on the r.h.s. are $K(\E)$ objects, i.e. they transform as
\be 
\cQ \longrightarrow k^{-1}(\pa_t + \cQ) k \qquad
\cP \longrightarrow k^{-1}\cP k \
\ee
Thus $\cQ$ plays the role of a $\kzehn$ gauge connection; however,
its full significance will become apparent only in a supersymmetric 
extension of the theory, where $\cQ$ is the quantity through which 
the bosonic degrees of freedom couple to the fermions (supposed to 
belong to a {\em spinorial} representation of $\KE$). The $\zehn$-invariant 
bosonic Lagrangian is the standard one for a point particle moving 
on the coset manifold $\EKE$:
\be\label{Lag}
\cL \equiv \cL(n,A,\pa_t A) := \ft12 n^{-1} \langle \cP| \cP \rangle
\ee
where $\langle .|.\rangle$ is the standard invariant bilinear form 
\Ref{Bil}. $n(t)$ is the lapse function required for the invariance
of the theory under reparametrizations of the time coordinate $t$, and whose 
variation yields the Hamiltonian constraint, which in turn ensures that 
the motion is along a null geodesic. Unlike finite dimensional simple Lie 
algebras, for which the number of independent polynomial Casimir invariants 
grows linearly with the rank, the bilinear form \Ref{Bil} is the {\em only 
polynomial invariant} for infinite dimensional KM algebras \cite{Kac}. 
For this reason, the Lagrangian \Ref{Lag} is {\em essentially unique}: 
its replacement by
\be\label{Lag'}
\cL' = n f \Big( n^{-2} \langle \cP| \cP \rangle \Big)
\ee
with $f(\xi) = \ft12 \xi + \cO (\xi^2)$ yields the {\em same} null
geodesic solutions as \Ref{Lag}. As a consequence {\it all couplings are 
already fixed by $\E$, and there is no need to invoke supersymmetry
or some other extraneous argument for this purpose.} For completeness 
we note that there are non-polynomial invariants \cite{KP}, which might 
be relevant for non-perturbative effects and the (conjectured) breaking 
of $E_{10}$ to $E_{10}(\Z)$, but these are not explicitly known.

From \Ref{Lag} we obtain the equation of motion
\be\label{EoM1}
n \cD (n^{-1} \cP) = 0
\ee
where $\cD$ denotes the $\KE$ covariant derivative whose action
%on any time-dependent Lie-algebra valued object $\cA$ 
is defined as
\be
\cD \cP := \pa_t \cP  + [ \cQ , \cP ]
\ee
Here we  omit the subscript $t$ on the covariant derivative $\cD$, and we
omit also to recall the fact that the covariant derivative  $\cD$ depends
on the solution of the geodesic equation we are writing down, through
its dependence on the value of $\cQ$. The simple looking compact form 
\Ref{EoM1} of the $\s$-model equations of motion is formally valid for 
any choice of gauge on the $\EKE$ coset space. On the other hand,
\Ref{EoM1} by itself does not constitute a well-defined, autonomous set 
of evolution equations. It must be completed by some (gauge-dependent) 
supplementary information telling us  how $\cP$ and $\cQ$ both depend 
on some basic coordinates, $A^{(\ell)}$, and their time derivatives.

The equations of motion \Ref{EoM1} are equivalent to the conservation 
of the $\E$ Noether charges
\be
\cJ = n^{-1} \cV \cP \cV^{-1}
\ee
which transform under rigid $\E$ as
\be
\cJ \longrightarrow g \cJ g^{-1} \qquad \mbox{for} \quad g\in\E
\ee

The main advantage of the triangular gauge is that \Ref{Lag} and 
\Ref{EoM1} are both well defined and computable if one analyzes the 
resulting equations level by level. We can thus expand (\ref{VdotV})
in non-negative levels according to 
\be
\cV^{-1} \pa_t \cV = Q^{(0)}\Dot L + P^{(0)}\Dot S + 
         P^{(1)}\Dot  E^{(1)} + P^{(2)} \Dot E^{(2)} + \dots
\ee 
Inspection shows that 
\be\label{PA}
P^{(\ell)} = \pa_t A^{(\ell)} + F_\ell \big(A^{(1)}, \pa_t A^{(1)},
        \dots , A^{(\ell - 1)}, \pa_t A^{(\ell -1)} \big)
\ee
where each $F_\ell$ is {\em polynomial} (of ascending order) and depends 
only on fields $A^{(n)}$ of {\em lower} level $n<\ell$.
With $F^{(1)} \equiv \big( E^{(1)} \big)^T$ {\it etc.} we next perform
the required split into compact and non-compact elements
\bea
\cQ &=& Q^{(0)}\Dot L + \frac12 P^{(1)}\Dot (E^{(1)} - F^{(1)})
            + \frac12 P^{(2)} \Dot (E^{(2)} - F^{(2)}) + \dots \nn\\
\cP &=& P^{(0)}\Dot S + \frac12 P^{(1)} \Dot (E^{(1)} + F^{(1)})
              + \frac12 P^{(2)} \Dot (E^{(2)} + F^{(2)}) + \dots
\eea
where, say, $Q^{(0)}\Dot L \equiv Q^{(0)}_{ab} L_{ab}$, with  
$L$ and $S$ from \Ref{Lab} and \Ref{Sab} above.

To write out the equations of motion we define a new `covariant derivative' 
operator, $\cD^{(0)}$, associated to rotations under the $SO(10)$ subgroup
by 
\be
\cD^{(0)}( P^{(0)}\Dot S) := \pa_t (P^{(0)}\Dot S) +
[Q^{(0)}\Dot L, P^{(0)}\Dot S]
\ee
when this operator acts on the level $\ell =0$ fields, and by
\bea\label{D0def}
\cD^{(0)} \big( P^{(\ell)} \Dot (E^{(\ell)} + F^{(\ell)}) \big)&:=&
\pa_t P^{(\ell)} (E^{(\ell)} + F^{(\ell)})  +  
[Q^{(0)} \Dot L\, ,\, P^{(\ell)} \Dot (E^{(\ell)} + F^{(\ell)})] \nn\\
&& - [P^{(0)}\Dot S \, ,\,  P^{(\ell)} \Dot (E^{(\ell)} - F^{(\ell)})] 
\eea
for $\ell \geq 1$. The second term on the r.h.s. is the expected
covariantization w.r.t. the $SO(10)$ subgroup, whereas the third term results
from the covariantization w.r.t. the remaining generators of $\kzehn$.
With this notation the equation of motion at level $\ell =0$ reads
\be\label{D0}
n \cD^{(0)} \big( n^{-1} P^{(0)}\Dot S\big) = - \frac12 \sum_{k=1}^\infty 
\Big[ P^{(k)}\Dot E^{(k)}\, ,\, P^{(k)} \Dot F^{(k)} \Big]
\ee
At levels $\ell\geq 1$ we similarly obtain
\bea\label{D0x}
n \cD^{(0)} \big( n^{-1} P^{(\ell)}
\Dot (E^{(\ell)} + F^{(\ell)})  \big)  &=&  - \sum_{k=1}^\infty
\Big[ P^{(\ell +k)}\Dot E^{(\ell +k)}\,,\, P^{(k)}\Dot F^{(k)}\Big]\nn\\
&& + \sum_{k=1}^\infty
\Big[ P^{(\ell +k)}\Dot F^{(\ell +k)}\, ,\, P^{(k)}\Dot E^{(k)} \Big]
\eea
The equations of motion thus consist of a derivative term and a term 
which is always quadratic in the `momenta' $P^{(k)}$. It takes only 
a little algebra to verify that \Ref{EoM1}, and hence \Ref{D0} and \Ref{D0x}
together are indeed equivalent to the standard geodesic equations 
on a coset manifold. Because we are here working with $\KE$ tensors, 
the `coordinates' $A^{(\ell)}$ do not appear explicitly in the above
equations.

\subsection{Equations of motion for levels $\ell\leq 3$}
To spell out the equations in more detail up to level $\ell =3$ we write
\bea\label{l3}
Q^{(0)}\Dot L &\equiv& Q^{(0)}_{ab} L_{ab} \quad , \qquad
P^{(0)}\Dot S \equiv P^{(0)}_{ab} S_{ab} \nn\\
P^{(1)}\Dot E^{(1)} &\equiv& \frac1{3!}\, P^{(1)}_{abc} E^{abc} \quad , \qquad
P^{(2)}\Dot E^{(2)} \equiv \frac1{6!}\,
P^{(2)}_{a_1 \dots a_6} E^{a_1 \dots a_6} \nn\\
P^{(3)}\Dot E^{(3)} &\equiv& \frac1{9!}\,
P^{(3)}_{a_0|a_1 \dots a_8} E^{a_0|a_1\dots a_8}
\eea
One easily checks that the covariant derivative $\cD^{(0)}$ introduced 
in \Ref{D0def} acts as
\be\label{D00}
\cD^{(0)} P^{(0)}_{ab}  = \pa_t   P^{(0)}_{ab} + 
Q^{(0)}_{ac}P^{(0)}_{cb} + Q^{(0)}_{bc}P^{(0)}_{ac} 
\ee
at level $\ell =0$, and like
\be\label{D0l}
\cD^{(0)} V_a = \partial_t V_a + Q^{(0)}_{ab} V_b - P^{(0)}_{ab} V_b
\ee 
on the vector indices of the higher level fields. We emphasize that 
all indices $a,b,\dots$ are to be treated as $SO(10)$ (`flat') indices, 
and therefore {\em all index contractions are performed with the 
flat metric $\d_{ab}$}. This property will be shown below to reflect
the fact that, under the $\E/\KE \leftrightarrow$ supergravity correspondence,
the $SO(10)$ subgroup of $\E$ can be identified with the $SO(10)$ subgroup of
the local Lorentz group $SO(1,10)$ in eleven dimensions. The link with
the anholonomic frames used in \cite{DaHeNi02} will be clarified 
in the following section.

For the comparison with the appropriately truncated equations of motion 
of $D\!=\! 11$ supergravity, we will impose the $\s$-model truncation
\be\label{trunc}
0 = P^{(4)} = P^{(5)} = P^{(6)} = \dots
\ee
To see that this is a {\em consistent truncation}, we note that each term 
on the r.h.s. of \Ref{D0x} contains a field of higher level than the 
l.h.s. of this equation. However, the vanishing of the infinite
tower of `momenta' $ P^{(\ell)}$ \Ref{trunc} does not imply the  
the vanishing or constancy of the associated infinite tower of 
coordinates $A^{(\ell)}$. Indeed, in view of \Ref{PA}, we find that  
the time evolution of the higher level component fields $A^{(\ell)}$ for 
$\ell\geq 4$ is generically non-trivial and determined (up to constants 
of integration) by the conditions \Ref{trunc}. Remarkably, the truncation 
to a finite number of low level `momenta' requires the excitation of the 
whole tower of $\E$ fields for its consistency!
Making use of the commutators for the first three levels (cf. section 2), 
we obtain 
\bea\label{EoMs0}
n \cD^{(0)} \big( n^{-1} P^{(0)}_{ab} \big) &=& 
- \frac14 \, \Big( P^{(1)}_{acd} P^{(1)}_{bcd} - 
     \frac19 \d_{ab} P^{(1)}_{cde} P^{(1)}_{cde}\Big)  \\
&-&  \frac1{2\cdot 5!} \, \Big( P^{(2)}_{ac_2\dots c_6} 
P^{(2)}_{bc_2\dots c_6} - 
\frac19 \d_{ab} P^{(2)}_{c_1\dots c_6} P^{(2)}_{c_1\dots c_6}\Big) \nn\\
&+& \frac4{9!} \,  
\Big( P^{(3)}_{c_1|c_2\dots c_8 a} P^{(3)}_{c_1|c_2\dots c_8 b}
    + \frac18 P^{(3)}_{a|c_1\dots c_8} P^{(3)}_{b|c_1\dots c_8} 
%\nn\\ && \qquad\qquad 
-\frac18  \d_{ab} P^{(3)}_{c_1|c_2\dots c_9} 
                    P^{(3)}_{c_1|c_2\dots c_9} \Big) \nn
\eea
at level $\ell =0$. At levels $\ell=1$ and $=2$, we have, respectively,
\bea\label{EoMs1}
n \cD^{(0)} \big( n^{-1} P^{(1)}_{abc} \big)&=& -\frac16\, 
     P^{(2)}_{abcdef} P^{(1)}_{def} + 
\frac{1}{3\cdot 5!}\,  P^{(3)}_{d_1|d_2\dots d_6abc} P^{(2)}_{d_1\dots d_6}  
\eea
and 
\bea\label{EoMs2}
n \cD^{(0)} \left( n^{-1} P^{(2)}_{a_1\dots a_6} \right) &=& 
\frac16 \, P^{(3)}_{b|cda_1\dots a_6} P^{(1)}_{bcd}
\eea
where higher level contributions have been suppressed in accordance 
with the cutoff \Ref{trunc}. Finally, at level $\ell =3$
\be\label{EoMs3}
n \cD^{(0)} \big( n^{-1} P^{(3)}_{a_0|a_1\dots a_8}\big) = 0
\ee 
More generally, it is easy to see that truncating at some higher level, 
the highest non-vanishing component of $\cP$ is always covariantly 
constant w.r.t. \Ref{D0def}.

\section{Comparison with $D=11$ supergravity}

We will now exhibit the relation between the $\s$-model equations
of motion derived in the foregoing section and the appropriately
truncated bosonic equations of motion of $D=11$ supergravity.
This relation can be obtained in two steps. First, we can formally
identify the objects $P$, $Q$ entering the compact, first-order form 
\Ref{EoMs0}-\Ref{EoMs3} of the $\s$-model equations of motion with some
corresponding objects entering the supergravity equations of motion
written in orthonormal frames. This preliminary identification will 
be explicitly performed in this section. However, because, as we said 
above, the equations \Ref{EoM1}  do not constitute an autonomous evolution 
system, it remains to check that the objects $P$, $Q$ defined in the 
first step, can indeed be derived from a consistent set of evolving 
coordinates $(A^{(\ell)} , \dot{ A^{(\ell)}})$ on the tangent bundle 
to the coset space $\E/\KE$. We will not explicitly perform this second 
step here, but show instead how the results of the first step match with
the previous results of \cite{DaHeNi02}. As the identifications obtained 
in \cite{DaHeNi02} were directly done for the autonomous second order 
form of the equations of motion, $ \ddot A = F(A, \dot A)$, the fact 
that our identifications for $\cP$, $\cQ$ can be matched to those 
of \cite{DaHeNi02} suffices to show that, indeed, $\cP$, $\cQ$ derive 
from a consistent set of $(A^{(\ell)} , \dot{ A^{(\ell)}})$ on the 
tangent bundle to the coset space $\E/\KE$.

The key point here is that we relate the time evolution of the
$\s$-model quantities, which depend only on the affine (time)
parameter $t$ to the time evolution of the spin connection and the 
field strengths and their first order spatial gradients {\em at an
arbitrary, but fixed spatial point} ${\bf x} = {\bf x}_0$. Taking into
account successively higher order spatial gradients will certainly require 
relaxing the cutoff conditions \Ref{trunc}. We do not know
at present how the matching works at higher levels.  We will
display below some of the terms which seem, within the present `dictionary',
problematic for the extension of the matching to higher levels.

For the bosonic equations of motion of  $D=11$ supergravity \cite{CJS} we
adopt the same conventions as in \cite{DaHeNi02}, except that we will
systematically project all quantities on a \textit{field of orthonormal 
frames}, {\it i.e.} on an  elfbein $E_M{}^A$. Here, $M,N,\cdots$ denote 
coordinate (world) indices, and $A,B,...$ flat indices  in $D=11$ with 
the metric $\eta^{AB}=(-+\cdots +)$. Using flat indices throughout, the 
supergravity equations of motion read
\bea\label{EoM11}
D_A F^{ABCD} &=& \frac1{8\cdot 144} \, \e^{BCDE_1...E_4 F_1...F_4}
                 F_{E_1...E_4} F_{F_1...F_4}\nn\\
R_{AB} &=& \frac1{12} \, F_{ACDE} {F_B}^{CDE} \, - \,
\frac1{144} \, \eta_{AB} F_{CDEF} F^{CDEF}
\eea
In addition we have the Bianchi identity 
\be\label{Bianchi11}
D_{[A} F_{BCDE]} = 0
\ee
Here $D_A$ is the Lorentz covariant derivative
\be
D_A V^B := \pa_A V^B + {{\o_{A}}^B}_C V^C
\ee
and the spin connection $ \o_{A\,BC} \equiv \eta_{BB'} {{\o_{A}}^{B'}}_C $ 
(which is antisymmetric in $BC$) is given by the standard formula
in terms of the coefficients of anholonomy $\Om_{ABC} $ (which 
are antisymmetric in the first pair of indices $AB$):
\be\label{omega}
\o_{A\,BC} = \frac12 \Big( \Om_{ABC} + \Om_{CAB} - \Om_{BCA} \Big)
\quad , \qquad
{\Om_{AB}}^C := {E_A}^M {E_B}^N \big( \pa_M {E_N}^C -\pa_N {E_M}^C \big) 
\ee
In flat indices, the Riemann tensor is
\be
R_{ABCD} = \partial_A \o_{B\, CD} - \partial_B \o_{A\, CD}  + 
{\Om_{AB}}^E \o_{E\, CD} + {\o_{A\, C}}^E \o_{B ED} - {\o_{B\, C}}^E \o_{A ED}
\ee
Next, we perform a 1+10 split of the elfbein, setting the shift $N^a=0$,
\be\label{elfbein}
E_M{}^A =
\left(\begin{array}{ccc}
          N &  0\\
          0 & e_m{}^a
\end{array}\right)
\ee
with the spatial zehnbein ${e_m}^a$. With this split, the coefficients of 
anholonomy become
\bea\label{Omega}
\Om_{abc} = 2{e_{[a}}^m {e_{b]}}^n \partial_m e_{nc} \; , \;\;
\Om_{0bc} = N^{-1} {e_b}^n \partial_t e_{nc} \; , \;\;
\Om_{a00} = \o_{0\, 0a} = - {e_a}^m N^{-1} \partial_m N
%   = {e_a}^m {e_b}^n \partial_m {e_n}^b
\eea
with all other coefficients of anholonomy vanishing.

The purely spatial components $\Om_{abc}$ can be separated into 
a trace  $\Om_a \equiv \Om_{abb} = \o_{bba}$ and a traceless part
$\tOm_{abc}$ (hence $\tOm_{abb}=0$)
\be
\Om_{abc} = \tOm_{abc} + \frac29 \Om_{[a} \d_{b]c}
\ee 
Below we will see that the respective equations of motion can 
only be matched if we set
\be\label{Oma}
\Om_a = 0
\ee
Because our analysis is local, {\it i.e.} takes place in some neighborhood 
of a given spatial point ${\bf x} = {\bf x}_0$, this condition can always 
be satisfied by a suitable choice of gauge for the spatial zehnbein.

Next we write out \Ref{EoM11} and \Ref{Bianchi11} with the (1+10) split of 
indices. To this aim, we separate the spatial components of the Ricci tensor 
in \Ref{EoM11} as
\be\label{Rab}
R_{ab} =: R_{ab}^{(0)} + R_{ab}^{(3)}
\ee
where the first term contains only time derivatives, and the second 
only spatial gradients (the superscripts are to indicate at which $\E$
levels these contributions become relevant). For the first term we 
obtain, with $\partial_0 \equiv N^{-1} \partial_t$, and remembering 
that $\eta^{00}=-1$ in our conventions,
\be\label{time}
R_{ab}^{(0)} = \partial_0 \o_{ab0} + \o_{cc0} \o_{ab0} - 2\o_{0c(a} \o_{b)c0} 
= N^{-1}e^{-1} \partial_t (eN^{-1} \o_{abt})  
  - 2 N^{-2} \o_{t \, c(a} \o_{b)\, ct} 
\ee
%&=& \nn\\
%&& \!\!\!\!\!\!\!\!\!\!\!\! \!\!\!\!\!\!\!\!\!\!\!\!\!\!\!\!\!\!
%\!\!\!\!\!\!\!\!\!\!\!\!\!\!\!\!\!\!\!\!\!\!\!\!
%N^{-2} \Big[ n \partial_t (n^{-1}\o_{abt}) + 
%2 \o_{t \, c(a} \o_{b)\, ct} \Big]
%\eea 
where $e \equiv {\rm det} \, {e_m}^a$. Recalling \Ref{D00} we see 
that $R^{(0)}_{ab}$ matches the structure of the l.h.s. of the 
$\ell =0$ equation of motion \Ref{EoMs0} up to an overall factor 
$N^{-2}$ if we equate
\be\label{ident0}
P^{(0)}_{ab}(t) = \o_{a\, b t} (t,{\bf x}) \big|_{{\bf x} = {\bf x}_0}
\qquad , \quad
Q^{(0)}_{ab} (t)= \o_{t\, ab} (t,{\bf x}) \big|_{{\bf x} = {\bf x}_0} 
\ee
where (cf. \Ref{omega} and \Ref{Omega})
\be\label{ident00}
\o_{t\, bc} = {e_{[b}}^n \partial_t e_{nc]} \quad ,\qquad
\o_{a\, bt} = {e_{(a}}^n \partial_t e_{nb)}
\ee
and identify the $\s$-model lapse function $n$ appearing in \Ref{Lag} 
and the lapse $N$ in \Ref{elfbein} via
\be\label{n}
n = Ne^{-1}  %      N=en  
\ee
(this quantity was called $\tilde N$ in \cite{DaHeNi03}). As we will
see more explicitly below this identification does not determine the 
time-independent (but space dependent) part of the spatial zehnbein, 
but it shows that the $\ell =0$ sector of the $\sigma$-model correctly 
reproduces the dimensional reduction of Einstein's equations to one 
time dimension.

As we will now show the remaining components in \Ref{l3} can be consistently 
related to the $D=11$ supergravity fields by the identification, or 
`dictionary', 
\bea\label{id}
P^{(1)}_{abc}(t) &=& F_{t \, abc} (t, {\bf x}) \big|_{{\bf x} = {\bf x}_0}
 \quad ( \equiv N F_{0abc} (t, {\bf x}) \big|_{{\bf x} = {\bf x}_0})   \nn\\
\quad P^{(2)}_{a_1\dots a_6}(t) &=& -\frac1{24}ne\e_{a_1 \dots a_6 bcde}
   F_{bcde}(t, {\bf x}) \big|_{{\bf x} = {\bf x}_0} \nn\\
P^{(3)}_{a_0|a_1 \dots a_8}(t) &=&\frac32 ne\e_{a_1 \dots a_8 bc} 
\tOm_{bc\,a_0} (t, {\bf x}) \big|_{{\bf x} = {\bf x}_0}
\eea
with $n$ from \Ref{n}.
This identification implies in particular that only the traceless part 
of $\Om_{abc}$ can appear on the r.h.s. of the formula for $P^{(3)}$ 
because \footnote{Inspection of the tables \cite{FisNic03} shows that 
also at higher levels there is no natural place for the trace $\Om_a$.}
\be 
P^{(3)}_{[a_0|a_1 \dots a_8]}=0 \quad \Longleftrightarrow \quad
\tOm_{ab\, b} = 0
\ee
The first two lines in \Ref{id} follow already by matching the 
r.h.s. of Einstein's equations \Ref{EoM11} with the $\ell=1,2$ terms 
on the r.h.s. of \Ref{EoMs0}. To see this we write out 
\bea
&& \!\!\!\!\! \!\!\!\!\!\!\!\!\!\! \!\!\!\!\!
\frac1{12} F_{aCDE} {F_b}^{CDE} - \frac1{144} \d_{ab} F^{CDEF}F_{CDEF}  \\
&=& - \frac14 F_{acd0} F_{bcd0}  + \frac1{36} \d_{ab} F_{cde0} F_{cde0} 
+ \frac1{12} F_{acde} F_{bcde}  - \frac1{144} \d_{ab} F_{cdef} F_{cdef}\nn
\eea
After multiplication with $N^2$ this agrees indeed with 
\be
 - \frac14 \Big( P^{(1)}_{acd} P^{(1)}_{bcd} - 
     \frac19 \d_{ab} P^{(1)}_{cde} P^{(1)}_{cde}\Big) 
 - \frac1{2\cdot 5!}\, \Big( P^{(2)}_{ac_2\dots c_6} 
P^{(2)}_{bc_2\dots c_6} - 
\frac19 \d_{ab} P^{(2)}_{c_1\dots c_6} P^{(2)}_{c_1\dots c_6}\Big) \nn
\ee
upon substitution of the $P^{(1)}$ and  $P^{(2)}$ from \Ref{id}.
The level $\ell =3$ contribution to Einstein's equation will be 
discussed below.

At this point the identifications at levels $\ell =1,2$ are fixed,
and the equations of motion and the Bianchi identity for $F_{ABCD}$
merely provide consistency checks on the identification \Ref{id}. 
The $abc$ component of the 3-form equation of motion yields
\be
D_0 F^{0abc} + D_e F^{eabc} =
- \frac1{144} \, \e^{abcd_1 d_2 d_3 e_1 \dots e_4} 
      F_{0 d_1 d_2 d_3} F_{e_1 \dots e_4}
\ee
Writing out the l.h.s. we obtain
\bea\label{lhs1}
D_0 F^{0abc}  + D_e F^{eabc} &=& 
-\partial_0 F_{0abc} - \o_{00e} F_{eabc} + 3 \o_{0e[a} F_{bc]0e} \nn\\
&&\!\! \!\! \!\!\!\!\!\!\!\!\!\!\!\!\!\!\!\!\!\!\!\!\!\!\!\!\!\!
\!\!\!\!\!\!\!\!\!\!\!\!\!\!\!\!\!\!\!\!\!\!\!\!\!\!\!\!\!\! 
+ \partial_e F_{eabc} - \o_{ee0} F_{0abc} + \o_{eef} F_{fabc}  
 -  3\o_{de[a} F_{bc]de} + 3 \o_{e 0[a} F_{bc]e0}
\eea
With the identification \Ref{n} and \Ref{id}, we get
\be
\partial_0 F_{0abc}+\o_{ee0} F_{0abc} + 3 \o_{0e[a} F_{bc]0e}
 - 3 \o_{e[a 0} F_{bc]e0}=  N^{-2}n \cD^{(0)} (n^{-1} P^{(0)}_{abc})
\ee
The remaining terms can be worked out to be
\bea 
\partial_e F_{eabc} + \o_{eef} F_{fabc} -  3\o_{de[a} F_{bc]de}
 - \o_{00e} F_{eabc} =  - \frac32 \tOm_{de[a} F_{bc]de} +
 N^{-1} \partial_e (N F_{eabc})
\eea
where we made use again of \Ref{Oma}. Relating the first term on 
the r.h.s. to the $P^{(3)} P^{(2)}$ term in \Ref{EoMs1} implies
the last formula in \Ref{id}. The second term $N^{-1}\partial_e (N F_{eabc})$
cannot be accounted for with the present truncation, and will require inclusion
of the higher level contributions. As mentioned in \cite{DaHeNi02} this
term formally corresponds to a term in the Hamiltonian which is of
`height' (with respect to the simple roots that it would contain as
exponent) higher or equal to  the level 30. A similar assertion holds
for the other spatial gradients that we shall neglect below.

To check the Bianchi identities \Ref{Bianchi11} with the $\ell=2$
equation \Ref{EoMs2} we write out
\bea
D_0 F_{abcd} + 4 \, D_{[a} F_{bcd]0} &=&
 \partial_0 F_{abcd} + 4 \o_{0e[a} F_{bcd]e} - 4 \o_{00[a} F_{bcd]0} \nn\\
&& \!\!\!\!\!\!\!\!\!\!\!\!\!\!\!\!\!\!\!\!\!\!\!\!\!\!\!\!\!\!
+ 4 \partial_{[a} F_{bcd]0} + 12 \o_{[ab \, e} F_{cd]e0} 
 + 4 \o_{[a \, 0e} F_{bcd]e}
\eea
As before, we recognize that
\be
\pa_0 F_{abcd}  + 4 \o_{0e[a} F_{bcd]e} + 4 \o_{[a \, 0e} F_{bcd]e}
= N^{-1} \cD^{(0)} F_{abcd}
\ee
Disregarding again space derivatives in the second term, as appropriate 
for our approximation, and using
\be
 12 {\o_{[ab}}^e F_{cd]e0} = 6 {\Om_{[ab}}^e F_{cd]e0}
\ee
together with the formula for $P^{(3)}$ from \Ref{id} we again obtain
perfect agreement.

Let us now return to the $\ell=3$ contributions. The corresponding terms
on the r.h.s. of \Ref{EoMs0} must be checked against the remaining
term in \Ref{Rab}, which is $(\pa_a \equiv {e_a}^m \pa_m$)
\be\label{Rab3}
R_{ab}^{(3)} = \frac14 \tOm_{cd\,a} \tOm_{cd\,b} 
- \frac12 \tOm_{acd} \tOm_{bc\,d}- \frac12 \tOm_{acd} \tOm_{bd\,c} 
- \frac12 \pa_c \tOm_{ca\,b} - \frac12 \pa_c \tOm_{cb\,a} 
\ee
where we have again dropped all trace terms in accordance with the gauge
choice \Ref{Oma}. On the other hand, substituting $P^{(3)}$ from 
\Ref{id} we get
\bea
&& \frac4{9!} \,\Big( P^{(3)}_{c_1|c_2\dots c_8 a} P^{(3)}_{c_1|c_2\dots c_8 b}
    + \frac18 P^{(3)}_{a|c_1\dots c_8} P^{(3)}_{b|c_1\dots c_8}
      -\frac18  \d_{ab} P^{(3)}_{c_1|c_2\dots c_9}
           P^{(3)}_{c_1|c_2\dots c_9} \Big) \nn\\
&& = \frac14 \tOm_{cd\,a} \tOm_{cd\,b}  - \frac12 \tOm_{ac\,d} \tOm_{bc\,d}
\eea
We thus see that the first two terms agree, but that the matching
fails for the other terms. However, as mentioned in \cite{DaHeNi02},
the terms that do not match correspond to second order spatial 
gradients, or to terms in the Hamiltonian that would involve the 
exponentiation of $\E$ roots of height 30 or more.

The final equation to be checked against \Ref{id} is \Ref{EoMs3}. 
With \Ref{ident0} and \Ref{ident00} we have
\be
\cD^{(0)} \big( e \e_{a_1 \dots a_{10}} \big) \propto
{e_c}^m \partial e_{m[c} \e_{a_1 \dots a_{10}]} = 0 
\ee
Noticing that $n$ drops out, and making use of \Ref{id} again, a little 
algebra shows that the $\ell =3$ equation of motion reduces to
\be\label{tOm}
\partial_t \tOm_{bc\,a} - \tOm_{bc\,d} {e_d}^m \partial_t e_{ma}
+ {e_b}^m \partial_t e_{md} \tOm_{dc\,a}
+ {e_c}^m \partial_t e_{md} \tOm_{bd\,a} = 0
\ee
Factorizing the spatial zehnbein as
\be\label{factor}
{e_m}^a (t,{\bf x}) = {\theta_m}^{\ba}({\bf x}) {S_{\ba}}^{a} (t)
\ee
the space dependent matrix ${\theta_m}^\ba$ drops out from the expressions 
for $Q^{(0)}$ and $P^{(0)}$ in \Ref{ident0} and \Ref{ident00}, and hence 
is left undetermined by the $\ell =0$ equation of motion. It is, however, 
fixed by the $\ell=1,2$ equations of motion, as we saw. Introducing the 
structure constants 
\be
{C_{\ba\bb}}^\bc := 2{\theta_{[\ba}}^m {\theta_{\bb]}}^n 
\partial_m {\theta_n}^\bc 
\ee
and substituting the ansatz \Ref{factor} into \Ref{tOm} we obtain
\be\label{Om}
\tOm_{abc} (t) = {(S^{-1})_a}^{\ba} (t) {(S^{-1})_b}^{\bb} (t) 
{C_{\ba\bb}}^{\bc} S_{\bc c} (t) \quad \Longrightarrow \quad
\pa_t {C_{\ba\bb}}^{\bc} = 0
\ee
This indeed solves \Ref{tOm} for constant and traceless ${C_{\ba\bb}}^{\bc}$.
We recall that, from \Ref{Rab3}, the matching at this level requires 
${C_{\ba\bb}}^{\bc}$ to be {\em spatially constant} as well.

In retrospect we recognize the factorization of the spatial zehnbein 
that was introduced in \cite{DaHeNi02}, following an earlier study of 
homogeneous cosmological solutions to $D=11$ supergravity in \cite{DHHS}. 
The background geometry is described by a purely spatial background frame 
\be\label{theta}
\theta^{\ba} = dx^m  {\theta_m}^{\ba}({\bf x})
\ee
whereas the time dependence of the zehnbein is entirely contained in the 
factor $S(t)$ and governed by \Ref{EoMs0}. Accordingly, in \cite{DaHeNi02}, 
all tensors
%\be
%d \theta^{\ba} = \ft12  {C_{\bb\bc}}^{\ba} \theta^{\bb}\wedge \theta^{\bc}
%\ee
were referred to the anholonomic frame $\theta^{\ba}$, and contracted 
with the purely time dependent metric
\be
g_{\ba\bb}(t) =  {S_{\ba}}^a (t){S_{\bb}}^a (t)
\ee
We can thus directly relate the $\s$-model fields used here and 
the quantities $\cD A$ used in \cite{DaHeNi02}; for instance,
\be
\cD A_{\ba\bb\bc} 
= {\theta_\ba}^m  { {\theta_\bb}^n \theta_\bc}^p \partial_t A_{mnp}
= {S_\ba}^a   {S_\bb}^b {S_\bc}^c P^{(1)}_{abc}
\ee
In contrast to the Lorentz tensors used here, the quantities
$\cD A_{\ba\bb\bc}$, {\it etc.} possess finite limits on the singular 
initial hypersurface, defining various `walls' as explained in 
\cite{DaHeNi03}. The frame metric $g_{\ba\bb} (t)$, on the other hand,
has no limit, but exhibits a singular behavior with chaotic 
oscillations as $t\rightarrow 0$. 

We thus see that the truncated $\s$-model equations of motion imply
the factorization on which the analysis of \cite{DaHeNi02} was based. 
Furthermore, the matching up to level $\ell=3$ with the cutoff \Ref{trunc} 
restricts the spatial geometry to frames with constant ${C_{\ba\bb}}^{\bc}$. 
Neither of these statements remains true if we relax \Ref{trunc}. 
For instance, including level four, the $\ell=3$ equations become, 
schematically,
\be
\cD^{(0)} P^{(3)} \sim P^{(4)} P^{(1)} \neq 0
\ee
with similar corrections for the $\ell < 3$ equations. Therefore,
the split into a space-dependent background frame, and a purely time 
dependent part $S$ no longer works. Moreover, when switching on 
higher levels, we expect that we will have to modify the identifications 
\Ref{id} which were found to work when only the first three levels were 
turned on; in other words, the `dictionary' is probably sensitive to the 
level at which we truncate. The challenge is now to find the (spatially) 
non-local and level-dependent correspondence between supergravity 
objects and $\s$-model ones, that will resolve the remaining discrepancies 
between $D=11$ supergravity and the $\E/\KE$ $\s$-model.

\subsection{Acknowledgments}
H.N. would like to thank the organizers for the invitation to a very
pleasant meeting. We are grateful to A.~Kleinschmidt for discussions 
and comments on the manuscript.

\end{document}